\documentclass{webofc}
\usepackage[varg]{txfonts}   
\usepackage{extarrows}
\usepackage[colorlinks=true,linkcolor=black]{hyperref}
\newcommand{\pcs}{P_{cs}(4459)}
\newcommand{\dvvb}{\bar{D}^{*}}
\newcommand{\db}{\bar{D}}

\newcommand{\jpsi}{J/\psi}

\newcommand{\dpn}{{D}^{0}}
\newcommand{\dvn}{D^{*0}}
\newcommand{\ds}{D_{s}^{-}}
\newcommand{\dsv}{D_{s}^{*-}}
\newcommand{\zca}{X(4020)}
\newcommand{\zc}{Z_c(3900)}
\newcommand{\zcs}{Z_{cs}(3985)}
\newcommand{\xic}{\Xi_c}
\newcommand{\xicp}{{\Xi'}_c}
\newcommand{\nn}{\nonumber}

\begin{document}
\title{Compositeness and several applications to exotic hadronic states with heavy quarks}
%
%

\author{\firstname{Jos\'e Antonio} \lastname{Oller}\inst{1}\fnsep\thanks{\email{oller@um.es}}
  \and
        \firstname{Zhi-Hui} \lastname{Guo}\inst{2}\fnsep\thanks{\email{zhguo@hebtu.edu.cn}}
}

\institute{Departamento de F\'{\i}sica, Universidad de Murcia, 30071 Murcia, Spain
  \and
Department of Physics,  Hebei Normal University,  Shijiazhuang 050024, China
}

\abstract{Several methods for studying the nature  of a resonance are applied to   resonances  recently discovered in the bottonomium and charmonium sectors.  We employ the effective-range expansion, the saturation of the  width and compositeness of a resonance, as well as direct fits to data. The latter stem from generic $S$-matrix parameterization that account for relevant dynamical features associated to channels that couple strongly in an energy region around the resonance masses, in which their thresholds also lie. We report on results obtained with these methods for the resonances $Z_b(10610)$, $Z_b(10650)$, $Z_{cs}(3985)$, $Z_c(3900)$, $X(4020)$, $X(6900)$, $X(6825)$, and  $P_{cs}(4459)$.}
\maketitle
\section{Introduction}

We use the methods reviewed in Ref.~\cite{oller2.theseproceedings}, 
and more details can be found there. In the subsequent we denote  this reference by [I], and we just write here the basic formulas that we are going to use. The compositeness  $X$ from the effective-range expansion (ERE) is calculated as $X={k_i}/{k_r}$, with the resonance momentum $k_R=\sqrt{2\mu (E_R-m_{\rm th})}=k_r-ik_i$ calculated in the second Riemann sheet (RS), so that $k_r\,,k_i\geq 0$. $E_R=M_R-i\Gamma/2$ is the resonance pole position with mass $M_R>m_{\rm th}$ ($m_{\rm th}$ is the threshold  of the active channel) and width $\Gamma$.

Beyond the ERE we also use the following formula for calculating  the partial compositeness coefficient $X_i$ of the $i_{\rm th}$ channel with its threshold above the resonance mass,
\begin{align}
  \label{221027.3}
  X_i&=\left|g_i^2 \frac{dG_i^{II}(s_R)}{ds}\right|~.
\end{align}
The superscript ${II}$ on the unitarity loop function $G_i(s)$  indicates that it is calculated in the second RS for those channels in which the transit to this sheet is involved to find the pole at $s_R$, otherwise it remains in the first RS ($s$ is the standard the Mandelstam variable). The function $G_i(s)$ is given in [I]. The couplings $g_i$ are calculated from the residues of the  partial-wave amplitudes (PWAs)  at the resonance pole position. In terms of them the partial decay width $\Gamma_i$  is calculated from two equations, depending on whether the threshold of the $i_{\rm th}$ channel is relatively far or close to  $M_R$ compared to its width $\Gamma$, respectively, as
\begin{align}
    \Gamma_i&=\frac{|g_i|^2}{8\pi M_R^2}~,~
\text{ or }~ \Gamma_i=|g_i|^2 \int_{m_{\rm th, i}}^{M_R+n\Gamma_R} dw \,\frac{q_i(w^2)}{16\pi^2 \,w^2} \frac{\Gamma}{(M_R-w)^2+\Gamma^2/4}~,~n\gg 1~, 
\end{align}
with  $q_i$ the center of mass momentum of the $i_{\rm th}$ channel.

\section{Resonances $Z_b(10610)$ and $Z_b(10650)$}
\label{sec.221022.2}

  We review the application performed in Ref.~\cite{Kang:2016ezb} of calculating  $X$ within the ERE to study the nature of the bottonomium resonances  $Z_b(10610)$ and $Z_b(10650)$ (also called $Z_b$ and $Z_b'$, in this order).  These are, respectively, $B^{(*)}\bar{B}^*$ systems with $I^G(J^P)=1^+(1^+)$ discovered in Ref.~\cite{Belle:2011aa} by the Belle Collaboration. The mass and width of the resonances determined in this reference are
\begin{align}
  \label{221022.5}
E_{Z_b}&=10607.2\pm 2.0-i(9.2\pm 1.2)~{\rm MeV}~,\\
E_{Z'_b}&=10652.2\pm 1.5-i(5.5\pm 1.1)~{\rm MeV}~,\nn
\end{align}
with the resonance mass below the corresponding $B^{(*)}\bar{B}^{*}$ threshold by around  3~MeV.  The resulting values for the scattering length $a$, effective range $r$, and the  compositeness $X$  are given in Table~\ref{tab.221022.1}. 

\begin{table}
\centering
\caption{\small Summary of scattering and compositeness properties of the $Z_b$ and $Z_b'$. \label{tab.221022.1}}
\begin{tabular}{lll}
\hline
               & $Z_b(10610)$ &  $Z_b(10650)$ \\
\hline
$a\,(\rm{fm})$ & $-1.03 \pm 0.17$  & $-1.18 \pm 0.26$\\
$r\,(\rm{fm})$ & $-1.49 \pm 0.20$ & $-2.03 \pm 0.38$ \\
$X$ & $0.75\pm 0.15$  & $0.67\pm 0.16$\\
\hline
\end{tabular}
\end{table}

For the $Z_b$,~$Z_b'$ the method based on saturating the total decay width gives  results that are almost identical to those in Table~\ref{tab.221022.1}, which is not surprising since the ERE method assumes only one channel. However, when saturating the partial decay width of the $Z_b^{(')}$ into $B^{(*)}\bar{B}^*$,  measured too in Ref.~\cite{Belle:2011aa}, one can account for coupled channels and the resulting compositeness coefficients for the $Z_b$ and $Z_b'$ are, respectively,
\begin{align}
X_{Z_b}&= 0.66\pm 0.11 ~,~~X_{Z_b'}= 0.51\pm 0.10~.
  \end{align}
The numbers are still compatible with Table 1, but they are on the lower side due to the saturation of the partial decay widths only.

\section{Study of the nature of the $Z_{cs}(3985)$, $Z_c(3900)$, and $X(4020)$}
\label{sec.221022.3}

We now report here on the unified study of Ref.~\cite{Guo:2020vmu} on the nature of the charmonium resonances $Z_{cs}(3985)$, $Z_c(3900)$, and $X(4020)$. The lighter  and heavier channels (the one with  its threshold close to the resonance mass) used by us for the study of every resonance are
\begin{align}
  \label{221022.6a}
&  Z_c (3900): J/\psi \pi,\,\bar{D}D^*/D\bar{D}^* (3875.5)\,;\,X(4020):h_c\pi,\, D^*\bar{D^*} (4017.1)\,;\\
 & Z_{cs}(3985): J/\psi K^-,~ D_s^-D^{*0}(3975.2)/D_s^{*-}D^0 (3977.0)\,.\nn
\end{align}
The thresholds for the heavier channels are given between parentheses in MeV.

\begin{table}
\centering
\caption{\small Values of $a$, $r$ and  $X$  from the elastic ERE study.  For the $\zcs$ the analysis is done twice by taking either the threshold of $\ds\dvn$ or $\dsv\dpn$. \label{tab.221022.3}}       
\begin{tabular}{ c c  c c c c}
\hline
 Charmonium & Mass   & Width &  $a$     & $r$   &  $X$  \\
  Resonance         & (MeV)  &   (MeV)    & (fm)    &  (fm) &   
\\ \hline
$\zc$  & $3888.4\pm 2.5$ & $28.3\pm 2.5$  & $-0.84 \pm 0.13$ & $-2.52 \pm 0.25$  & $0.45\pm 0.06$
\\ 
$\zca$ & $4024.1\pm 1.9$ & $13\pm 5$    & $-1.04 \pm 0.30$ & $-3.90 \pm 1.35$  & $0.39\pm 0.14$ 
\\ 
$\zcs$ & $3982.5\pm 3.3$ & $12.8\pm 6.1$  & $-1.00 \pm 0.47$ & $-4.04 \pm 1.82$ & $0.38\pm 0.18$ \\
       &   &                             & $-1.28 \pm 0.60$ & $-3.65 \pm 1.60$ & $0.46\pm 0.19$ 
\\\hline
\end{tabular}
\end{table}

The elastic ERE study, which considers only each channel with its threshold around the resonance mass, yields the results shown in Table~\ref{tab.221022.3}.
 One can observe from Table~\ref{tab.221022.3} several interesting points. Firstly, it is notorious that the values for $a$, $r$ and $X$ are quite similar among all the three resonances. This is a clear hint towards a remarkable similar dynamics in the constitution of all these resonances (a suspicion that one could already have considering the heavier channels shown in Eq.~\eqref{221022.6a}, being related by heavy-quark spin symmetry and $SU(3)$). Secondly, we note the rather large and negative values of $r$ and that the compositeness is less than 0.5 for all the resonances. These two aspects are related as discussed in Sec.~7 of [I].

Next, the method based on the saturation of the width and compositeness of every resonance is also considered. This method fixes the two couplings in the case of the resonance $Z_c(3900)$, because its separate decay widths into the two channels in Eq.~\eqref{221022.6a} have been measured \cite{BESIII:2013qmu}. This analysis gives
\begin{align}
\label{221022.6}
Z_c(3900):&~\frac{\Gamma_{D\bar{D}^*}}{\Gamma_{J/\psi\pi}}=6.2\pm 2.9\,,\,|g_1|= 1.46_{-0.23}^{+0.43}\,,  |g_2|= 7.89_{-0.44}^{+0.18}\,,\,|g_1|\ll |g_2|~,\\
&X_1=0.002\pm0.001\,,\, X_2= 0.436_{-0.047}^{+0.021}\,,\,X=X_1+X_2= 0.438_{-0.047}^{+0.021}\,.\nn 
\end{align}
It is worth stressing that  $X$ is almost identical to the one determined in Table~\ref{tab.221022.3} from the ERE method.
Then, given the analogous dynamics of the constituents of these resonances, as reflected in Table~\ref{tab.221022.3}, 
we then take as the compositeness $X$ for the $Z_{cs}(3985)$ and $X(4020)$ the one determined from ERE study.
In this way we can solve for the couplings and predict the partial-decay widths of these resonances. The results of the analysis are given in Table~\ref{tab.221022.4}.

\begin{table}
  \centering
  \caption{\small  The coupled-channel solutions for the $\zca$ and $\zcs$. \label{tab.221022.4}} 
\begin{tabular}{ c c c c c c c }
\hline
Charmonium  & $|g_1|$ & $|g_2|$   & $\Gamma_1$ & $\Gamma_2$  & $X_1\times 10^{3}$   & $X_2$    \\
Resonance       &  (GeV)  & (GeV)     & (MeV)     &  (MeV)      &         &        \\
\hline
$\zca$    &   &      &      &        &         &        \\
$X_\text{ERE}=0.39\pm 0.14$ & $1.1 \pm 0.2 $ & $6.5\pm 1.3$ & $1.4\pm 0.5$  & $11.6\pm 4.5$ & $1\pm 1$ &  $0.39\pm 0.14$\\
$\zcs$ &   &    &  &   &   &   \\
Threshold$({\ds\dvn})$ &  &      &     &       &         &     \\
$X_{\text{ERE}}=0.38\pm 0.18$ & $0.8\pm 0.2$ & $6.4\pm 1.7$ & $1.2\pm 0.6$  & $11.6\pm 5.3$ & $0.8\pm 0.4$ &  $0.38\pm 0.18$\\
Threshold$({\dsv\dpn})$ &  &      &     &       &         &      \\
$X_{\text{ERE}}=0.46\pm 0.19$  & $0.9\pm 0.2$ & $6.8\pm 1.7$ & $1.2\pm 0.6$  & $11.6\pm 5.6$ & $0.8\pm 0.4$ &  $0.46\pm 0.19$\\
\hline
\end{tabular}
\end{table}

\section{$X(6900)$ and $X(6825)$}
\label{sec.221022.4}

\begin{figure}
    \centering
    \sidecaption
    \includegraphics[width=0.5\textwidth,angle=0]{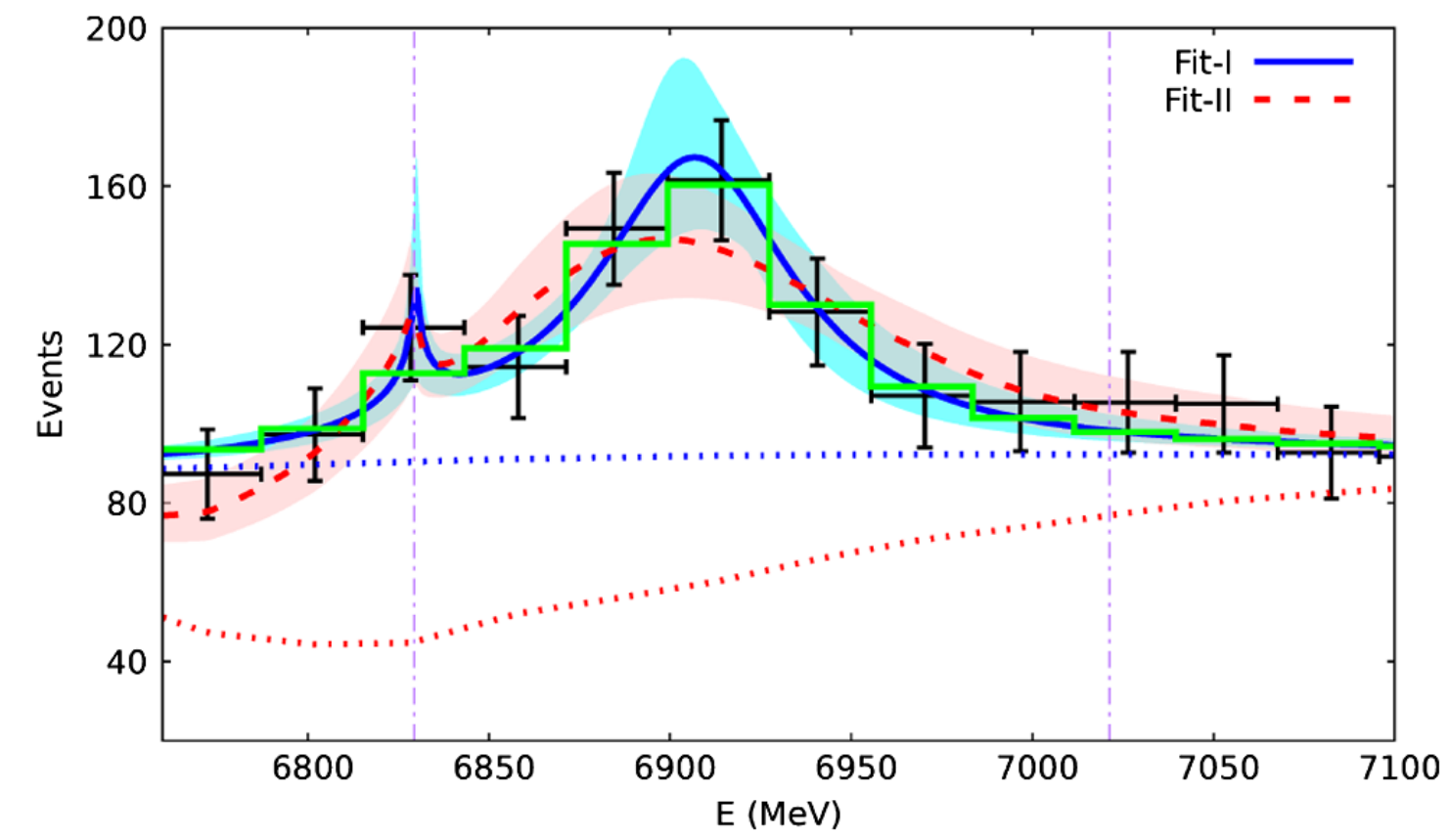}
    \caption{\small $J/\psi J/\psi$ event distribution around the resonant region of the $X(6900)$.\label{fig.221022.1}}
    \end{figure}

The LHCb Collaboration \cite{LHCb:2020bwg} discovered the  fully charmed tetraquark resonance $X(6900)$ in the di-$\jpsi$  mass distributions. The fitted mass and width for this resonance are,
\begin{align}
  \label{221027.1}
  \text{Model I:}~ M&=6905\pm 11\pm 7~\text{MeV}\,,~ \Gamma=80\pm 19\pm 33~\text{MeV}\\
  \text{Model II:}~M&=6886\pm 11\pm 11~\text{MeV}\,,~\Gamma=168\pm 33\pm 69~\text{MeV}~,\nn
  \end{align}
  depending on the model used for the non-resonant signal. Here we review the approach and results of Ref.~\cite{Guo:2020pvt}, which studied the $S$-wave scattering in coupled channels of $J/\psi J/\psi$($\eta_c\eta_c$), $\chi_{c0}\chi_{c0}(6829.4)$, and $\chi_{c1}\chi_{c1}(7021.3)$ around the thresholds (given between brackets in MeV) of the latter two channels. The symbol  $J/\psi J/\psi$($\eta_c\eta_c$) refers to the fact that by studying explicitly the $\jpsi\jpsi$ channel one is also accounting for other light channels like the $\eta_c\eta_c$, whose thresholds are far away from the mass of the $X(6900)$.  
  The $T$ matrix is calculated with a unitarization formula  \cite{Oller:2019opk,Oller:2019rej} that considers  linear interactions in $s$ between the channels  $\chi_{c0}\chi_{c0}$ (2) and $\chi_{c1}\chi_{c1}$(3), allowing for a CDD pole at $M_{CDD}^2$. The relevant formulas are
\begin{align}
\label{221022.11}
\mathcal{T}(s) = [1-\mathcal{V}(s)\cdot G(s)]^{-1} \cdot \mathcal{V}(s)\,;\,\mathcal{V}(s)= \left(\begin{matrix}
      0  &  b_{12}  & b_{13} \\
      b_{12} &  \frac{b_{22}}{M_{\jpsi}^2} (s- M_{CDD}^2) & \frac{b_{23}}{M_{\jpsi}^2} (s- M_{CDD}^2)\\
      b_{13} &  \frac{b_{23}}{M_{\jpsi}^2} (s- M_{CDD}^2) & \frac{b_{33}}{M_{\jpsi}^2} (s- M_{CDD}^2)\\
   \end{matrix}\right) \,.
\end{align}
Because of the heavy-quark spin symmetry (HQSS)  one has that $b_{13}=\frac{b_{12}}{\sqrt{3}}\,$, $b_{23}=\frac{b_{22}}{\sqrt{3}}\,,$ and $b_{33}=\frac{b_{22}}{3}$~. The $G(s)$ matrix is diagonal and it contains the unitarity loop functions $G_i(s)$ [I]. The subtraction constant $a(\mu)$ is estimated by matching at threshold $G_i(s)$ with its value when calculated with a three-momentum cutoff $\mu$ around 1 GeV. 
 In order to reproduce the di-$\jpsi$ event distribution we take into account the final-state interactions due to the interaction among the three channels with the generic formula \cite{Oller:2019opk,Oller:2000vu,Oller:1999ag}
  \begin{align}
\label{221022.12}
B(s) & = [1-\mathcal{V}(s)\cdot G(s)]^{-1}\cdot \mathcal{P}~~;~~\mathcal{P} = \left(\begin{matrix}
                  0 \\
                  d_{2} \\
                  d_{2}/\sqrt{3} 
                 \end{matrix} \right) ~,
  \end{align}
  where $\mathcal{P}$ is a vector containing the production vertices $d_i$, with $d_1=0$ due to the assumed weak coupling of the $\jpsi\jpsi$. We have also checked that fits are stable if releasing $d_1$. Note also that in Eq.~\eqref{221022.12} we have employed HQSS to set $d_3=d_2/\sqrt{3}$. The final formula for calculating the di-$\jpsi$ event distribution is 
\begin{align}
\label{221022.13}
\frac{ d \mathcal{N}(s)} {d \sqrt{s}} = |B_1(s)|^2 \frac{q_{{\jpsi}{\jpsi}}(s)}{M_{\jpsi}^2}~,
\end{align}
in obvious notation. The free parameters in our approach are then  $b_{12}$, $b_{22}$, $M_{CDD}^2$, and $d_2$, which are fitted to data.  We distinguish between the Fits I and II corresponding to the non-resonant background being treated as in the Models I and II of Ref.~\cite{LHCb:2020bwg}, respectively. The fitted values for the free parameters are given in Table~\ref{tab.221022.6}.  The reproduction of the data (crosses) by the fits around the resonance is shown in Fig.~\ref{fig.221022.1} by the solid (Fit-I) and dashed (Fit-II) lines. The dotted lines show the non-resonant backgrounds taken from Ref.~\cite{LHCb:2020bwg}, and the green histogram is the average of the Fit-I  over the experimental bin width of 27~MeV. 
 The resulting $|g_i|$ and $X_i$ are given in Table~\ref{tab.221022.7}. We notice that the mass and width of the resonance  are well compatible with those determined by the LHCb \cite{LHCb:2020bwg}, cf. Eq.~\eqref{221027.1}.  The resulting total compositeness is quite small, $X<0.2$, which is a clear hint of a large bare ``elementary'' component of the $X(6900)$.
This is in agreement with the fact of having the CDD pole so close to the resonance mass [I], $M_{\rm CDD}\approx M_R$, which is also tightly connected with the Morgan's pole-counting criterion \cite{Morgan:1990ct,Morgan:1992ge}. We have checked that the pole also appears in the 5th RS $(-,-,-)$ (for the definition of RSs see [I]).  We show in Ref.~\cite{Guo:2020pvt} that a way to distinguish between  Fits I and II could be measuring the invariant mass distributions of $\chi_{c0}\chi_{c0}$. 

\begin{table}
\centering
\caption{\small    Fits I and II. The entries marked with asterisks are fixed during the fits. \label{tab.221022.6}}
\begin{tabular}{ c c c c c c c c c }
\\ \hline
& $\chi^2/{\rm d.o.f}$ & $a(\mu)$ & $M_{CDD}(\text{MeV})$ & $b_{22}$  & $b_{12}$  & $d_2$   
\\ \hline 
Fit-I  & $1.6/(12-3)$ & $-3.0^{*}$ & $6910^*$   & $10817_{-2096}^{+8378}$  & $151_{-99}^{+153}$  & $2213_{-316}^{+2106}$  \\  
Fit-II & $4.9/(12-3)$ & $-3.0^{*}$ & $6885^*$   & $21085_{-7359}^{+15141}$ & $484_{-112}^{+239}$  & $3646_{-714}^{+1325}$  \\  \hline
\end{tabular}
\end{table}

\begin{table}
\centering
\caption{\small  Resonance poles of the $X(6900)$ and its properties. \label{tab.221022.7}}
\begin{tabular}{ c c c c c c c c }\hline
 & Mass & Width/2 & $|g_1|$ & $|g_2|$ & $|g_3|$ & $X=\sum_{i=1}^{3} X_i$   
\\
&(MeV)  & (MeV)     & (GeV)      &  (GeV)    &   (GeV)  \\
\hline
 Fit-I   & $6907_{-3}^{+5}$ & $33_{-10}^{+14}$ & $4.6_{-2.8}^{+2.5}$  & $9.7_{-2.6}^{+1.4}$ & $5.6_{-1.5}^{+0.8}$  &  $0.17_{-0.07}^{+0.04}$  \\
 Fit-II  & $6892_{-2}^{+2}$ & $80_{-17}^{+24}$ & $10.3_{-1.4}^{+1.8}$  & $6.9_{-1.9}^{+1.4}$ & $4.0_{-1.1}^{+0.8}$  &  $0.13_{-0.03}^{+0.03}$ 
\\ \hline
\end{tabular}
\end{table}

  
  The solid line in Fig.~\ref{fig.221022.1} shows a clear peak around 6825~MeV, which is washed out when taking the average over the bin width. This peak corresponds  to a pole in the 4th RS ($+,-,+)$ [I], a new resonance that we call the $X(6825)$. Its properties are summarized in Table~\ref{tab.221023.1}. Comparing this table with Table~\ref{tab.221022.7} for the $X(6900)$, we observe that the coupling to $\jpsi\jpsi$ of the former is much smaller, driving a much smaller width. At the same time, the couplings $|g'_2|$ and $|g'_3|$ are much larger than those of the $X(6900)$.  The new resonance is a $\chi_{c0}\chi_{c0}$ virtual state present only at the 4th RS [I]. According to Morgan's pole counting rule \cite{Morgan:1990ct,Morgan:1992ge} it should be of molecular nature.  

  \begin{table}
\centering
\caption{\small Pole position and couplings of the $X(6825)$. \label{tab.221023.1}}
\begin{tabular}{lllll}
  \hline
  Fit &   $E'_R$~MeV & $|g'_1|$ & $|g'_2|$ & $|g'_3|$ \\
  \hline
      I & $6827.0^{+1.6}_{-4.8}-i\,1.1^{+1.3}_{-1.0}$ & $1.4^{+0.6}_{-0.9}$ & $11.9^{+3.2}_{-3.1}$ & $6.8^{+1.8}_{-1.8}$ \\
      II & $6820.6^{+3.0}_{-2.7}-i\,4.0^{+1.7}_{-1.6}$ & $2.5^{+0.5}_{-0.6}$ & $15.8^{+0.7}_{-0.6}$ & $9.1^{+0.4}_{-0.4}$\\
      \hline
      \end{tabular}
\end{table}

  

We also investigated in \cite{Guo:2020pvt} the importance of the
channel $\psi(3770)J/\psi$, which threshold is at $6870.6$~MeV. By taking  the channels $\jpsi\jpsi$ (1) and $\psi(3770)J/\psi$ (2) the resulting fits  are not well determined because the fitted parameters are affected by huge errors. In addition, the coupling to $\psi(3770)J/\psi$ is much smaller than those to  $\chi_{c0}\chi_{c0}$ and  $\chi_{c1}\chi_{c1}$, a fact that clearly indicates a much less important role of the $\psi(3770)J/\psi$ channel. 
Related to these observations, Ref.~\cite{Guo:2020pvt} also considered a perturbative treatment of the $\psi(3770)J/\psi$ as the fourth channel, so that  this channel interacts only through its couplings to $\chi_{c0}\chi_{c0}$ and $\chi_{c1}\chi_{c1}$. 
It comes out that the Fits I and II obtained with only the first three channels are stable and the conclusions remain unchanged.


\section{Study of the $P_{cs}(4459)$}
\label{sec.221023.1}

Here we review the work of Ref.~\cite{Du:2021bgb} on the nature of the  $P_{cs}(4459)$ \cite{Wu:2010jy}, a charmonium pentaquark resonance with strangeness recently discovered by the LHCb Collaboration \cite{LHCb:2020jpq} in the $\jpsi \Lambda$ mass distribution. The mass and width of the resonance from this reference are given in Table~\ref{tab.221023.2}. 
  In Ref.~\cite{LHCb:2020jpq} the question whether the peak structure unveiled was due to one or two resonances with $J=1/2$ or $3/2$ was also raised. Reference~\cite{Du:2021bgb} applied three methods to study the nature of this resonance, and all of them clearly signal it is a $\Xi_c\bar{D}^*$ molecular resonance. We skip the discussions on the hidden-charm pentaquark states $P_c(4312)$, $P_c(4440)$ and $P_c(4457)$ in Ref.~\cite{Guo:2019kdc} due to  length limitations.

\begin{table}
\centering
\caption{\small Results of the $P_{cs}(4459)$ for $a$, $r$ and $X$ from the ERE method. The relevant channel  is indicated in the third column. \label{tab.221023.2}} 
\begin{tabular}{ c c c c c c}
\hline
 Mass   & Width & Threshold  & $a$     & $r$  & $X$    \\
  (MeV)  & (MeV) &  (MeV)    & (fm)    &  (fm)   
\\ \hline 
 $4458.8\pm 5.5$ & $17.3\pm 10.3$ & ${\Xi'}_c \bar{D}$~(4446.0)  & $-0.63\pm 0.38$ &  $-3.68\pm 2.11$ & $0.31\pm 0.19$  \\
 $4458.8\pm 5.5$ & $17.3\pm 10.3$ & $\Xi_c \bar{D}^*$~(4478.0)  & $-1.79\pm 0.23$ & $-0.94\pm 0.13$ & $--$
\\  \hline
\end{tabular}
\end{table}

First, the elastic ERE method  is separately applied to the channels $\xicp \bar{D}$ and $\xic \bar{D}^*$ whose thresholds are 4446.0~MeV and 4478.0~MeV, respectively. The results for $a$, $r$ and $X$ are given in Table~\ref{tab.221023.2}. 
 For the case of the $\xicp \bar{D}$ channel the value of $X$ is rather small, which is also consistent with having a large and negative $r$. That is, the resonance $P_{cs}(4459)$ has other more important components than the $\xicp \bar{D}$. On the other hand, when considering the case with the $\Xi_c D^*$ the effective range has a natural value for strong interactions, which is a hint towards a $\xic \bar{D}^*$  molecular nature of the resonance. 

  The consideration by Ref.~\cite{Du:2021bgb} of the method based on saturating $X$ and $\Gamma$ of the $\pcs$ for the coupled channels $\jpsi\Lambda$-$\Xi_c\bar{D}^*$, on the one hand, and $\jpsi\Lambda$-$\Xi_c'\bar{D}$, on the other, implies that no solution for $X\gtrsim 0.3$ is obtained for the latter case. Thus, the $\pcs$ cannot be a $\Xi_c'\bar{D}$  molecular-type resonance.  However, it could be a  $\Xi_c\bar{D}^*$ one since values of $X$ as large as 1 can be reproduced.

\begin{figure}
  \centering
  \begin{tabular}{ll}
  \includegraphics[width=.4\textwidth]{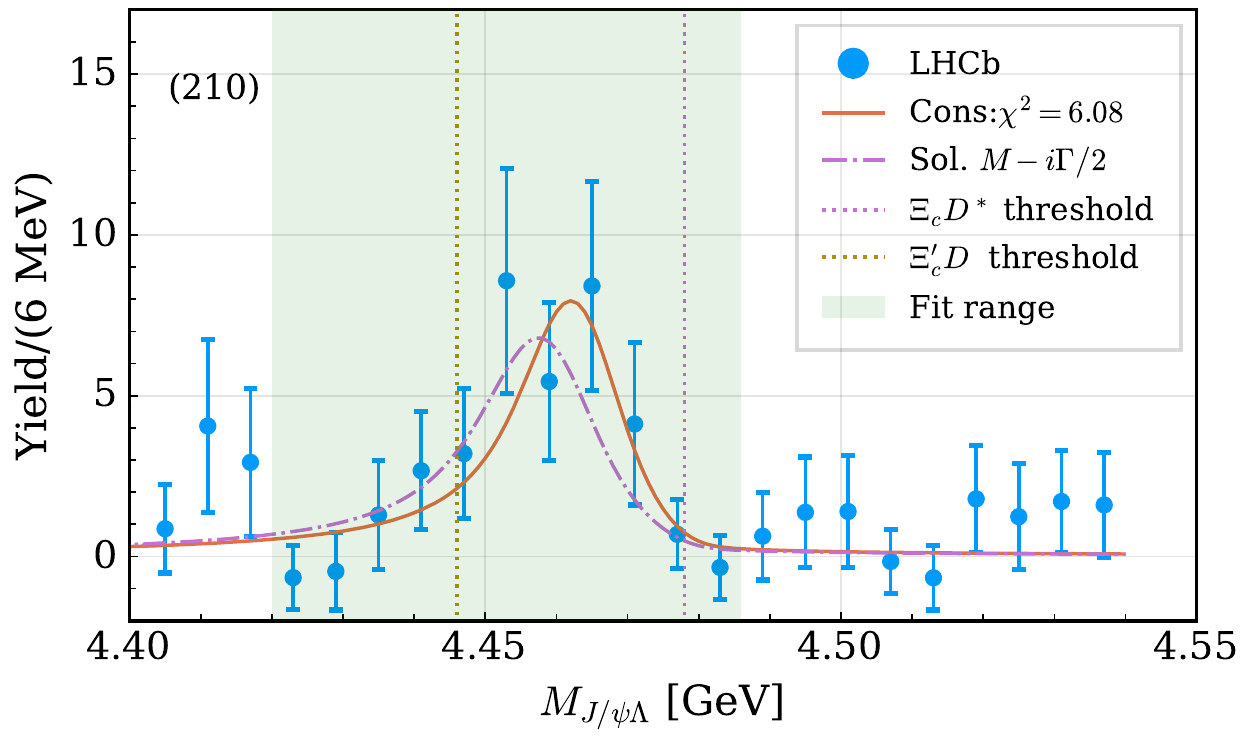} &
  \includegraphics[width=.4\textwidth]{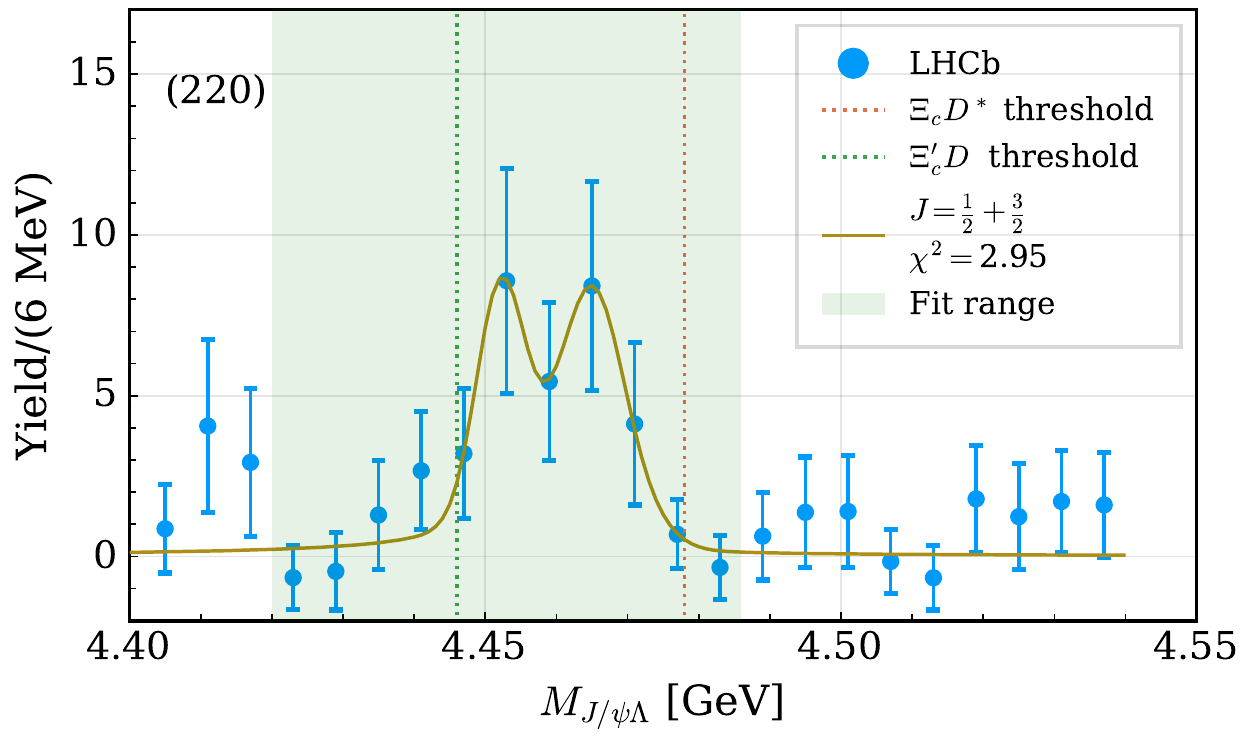} \\
  \includegraphics[width=.4\textwidth]{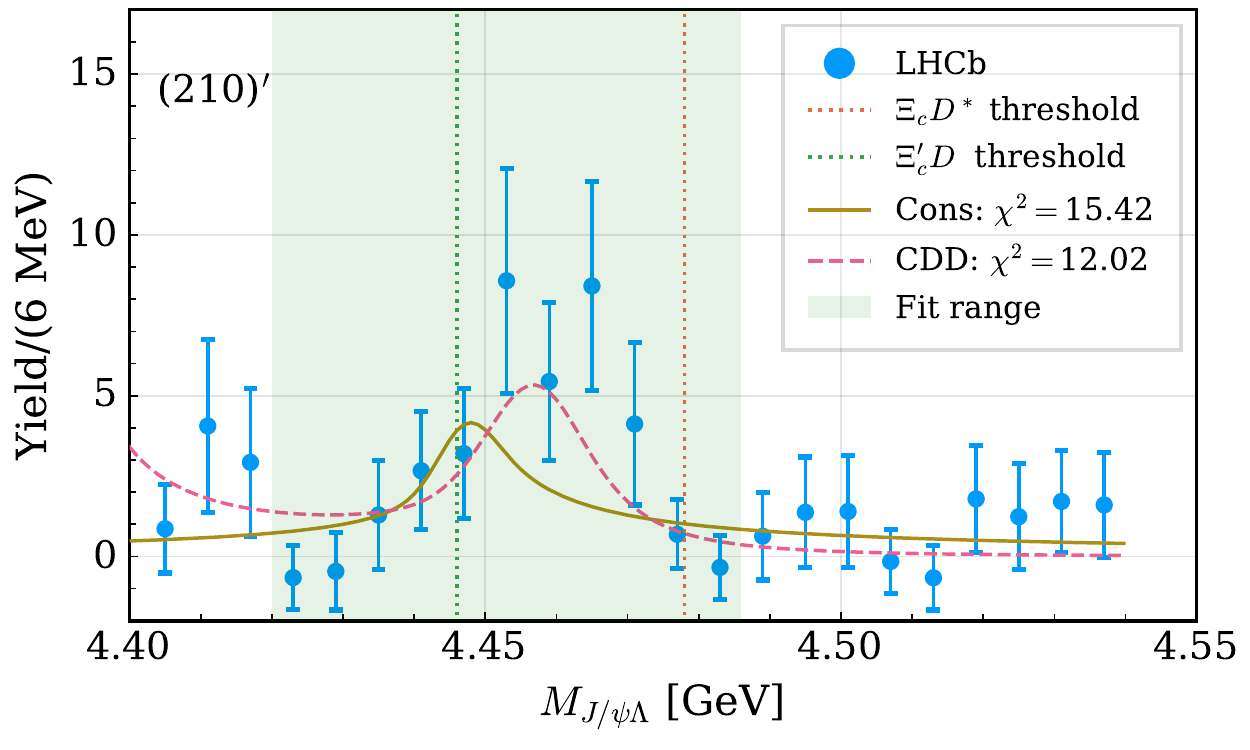} &
  \includegraphics[width=.4\textwidth]{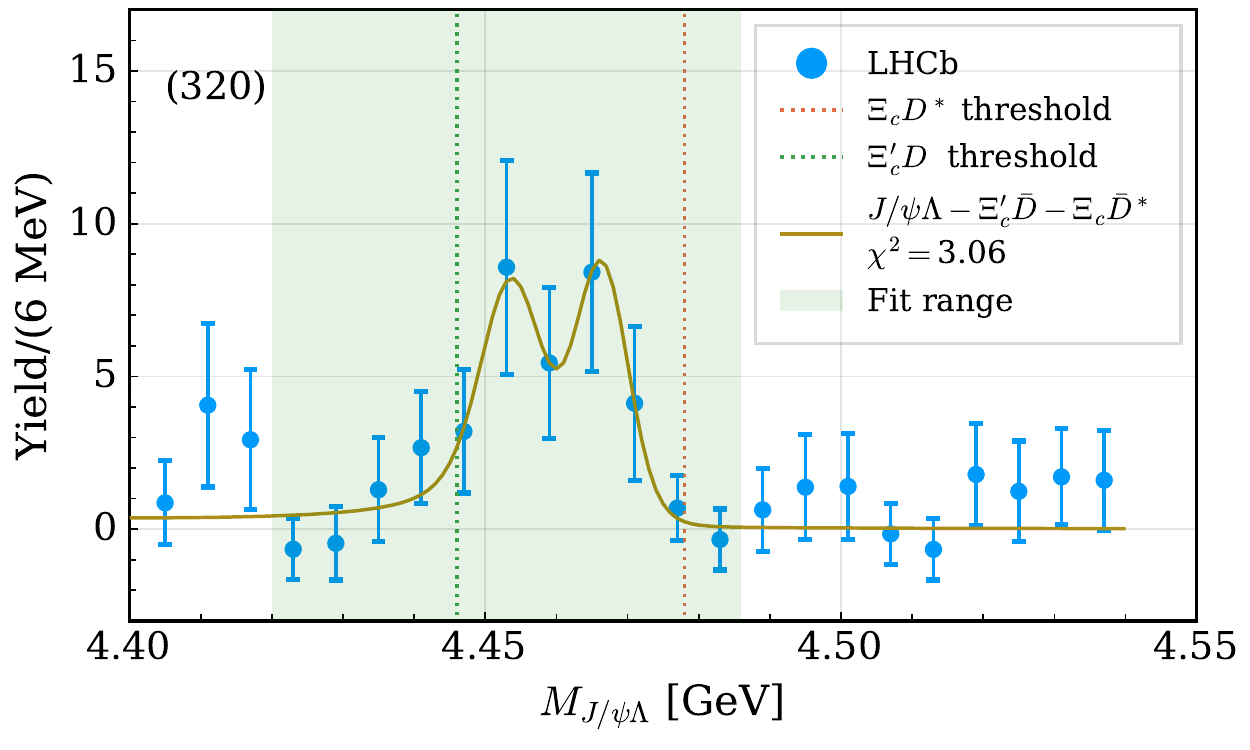}
  \end{tabular}
\caption{Reproduction of the experimental data by the different fits around the $P_{cs}(4459)$ resonance. The actual fitted area has a green shadow. The dot-dashed line curve in (210) results after imposing that $\Delta_{1/2}=0$ at the mass and width values of Ref.~\cite{LHCb:2020jpq}.   \label{fig.221023.1}}
    \end{figure}

The last method that Ref.~\cite{Du:2021bgb} applied  is to directly fit the data on the $\jpsi\Lambda$ event distribution provided by Ref.~\cite{LHCb:2020jpq}. Here, we consider a two coupled-channel problem involving the $\jpsi \Lambda$ and $\Xi_c\bar{D}^*$ channels interacting in $S$ wave. The inclusion of the channel $\eta_c\Lambda$ for $J=1/2$ would not add any new aspect \cite{Du:2021bgb}. 
 We restrict to $S$-wave scattering because the $P_{cs}(4459)$ is very close to the thresholds of $\Xi_c'\bar{D}$ and $\Xi_c\bar{D}^*$. It is also the case that in the HQSS  $J/\psi\Lambda$ ($\eta_c\Lambda$) cannot couple to the heavier channels considered in $D$ and higher partial waves. 
The basic formulas to calculate the PWAs in $S$ wave are similar to Eq.~\eqref{221022.11} with the interaction kernel ${\cal V}_J$, $J=1/2$ and $3/2$,  given by 
\begin{align}
\label{221023.15}
\mathcal{V}_{\frac{1}{2}}&=  \left(  \begin{array}{ll}
    0 &  g\\
 g & C_\frac{1}{2}
\end{array}\right)\,~~;~~
\mathcal{V}_\frac{3}{2}=\left(
\begin{array}{lll}
  0 & g \\
  g &   C_\frac{3}{2}
\end{array}
\right)~.
\end{align}
The HQSS requires that $C_\frac{1}{2}=C_\frac{1}{3}$, but we have let them to float in Eq.~\eqref{221023.15} as a check of the  completeness of the model. The $J/\psi \Lambda$ production amplitudes $F_J(s)$ and event distribution are given by 
\begin{align}
\label{221023.16}
  F_J(s)&=\frac{d_J}{\Delta_J(s)} 
  =\frac{d_J}{1-(C_J+G_1(s)g^2)G_3(s)}~~;~~  \Delta_J(s)={\rm det}\left[1-\mathcal{V}_J\cdot G(s)\right]~, \\
\frac{dN(s)}{d\sqrt{s}}&=\frac{1}{128\pi^3 M^3_{\Xi_b}}\frac{\sqrt{\lambda(M^2_{\Xi_b},s,M_K^2)\lambda(s,M^2_{\jpsi},M_\Lambda^2)}}{\sqrt{s}}\sum_{J}|F_J|^2~.\nn
\end{align}
 In addition, a convolution to take into account energy resolution of the experimental data is also made. 
 The results of the fits with this scheme based on using the two channels $\jpsi\Lambda$, $\Xi_c\bar{D}^*$ are given by the first two lines of Table~\ref{tab.221023.5}, and shown in Fig.~\ref{fig.221023.1}. We designate the fits by the tuple $(abc)$, where $a$ gives the number of channels, $b$ the number of partial wave amplitudes and $c$ the maximum degree in powers of $s$ of the matrix elements of ${\cal V}$. In the first line only the $J=1/2$ PWA is included (this is why $C_{3/2}=d_{3/2}=0$) and the fit is not good. In the second line both waves are included, the fit is good and has a much lower $\chi^2$, but the couplings $C_{1/2}$ and $C_{3/2}$ are very different, which indicates a clear violation of the HQSS expectation. Other fits were also explored attending to the possible presence of CDD poles by substituting $C_{1/2}$ with $C_{1/2}(s/M_{\rm CDD}^2-1)$.  However, the resulting fits are ruled out because they drive to poles in the first RS. As an informative remark, Ref.~\cite{Du:2021bgb} also studied in a similar way the coupled-channel system $\jpsi \Lambda$-$\Xi_c'\bar{D}$, which only interacts in $J=1/2$. This fit is denoted by $(210)'$ in Fig.~\ref{fig.221023.1} and is very poor. 

    \begin{table}
  \centering
  \caption{\small Fit parameters to data in the region of the $P_{cs}(4459)$. \label{tab.221023.5} }
  \begin{tabular}{lllllll}
    \hline
    Fit   & $\chi^2$ & $g$                 & $C_\frac{1}{2}$            & $d_\frac{1}{2}$            &  $C_\frac{3}{2}$ & $d_\frac{3}{2}$ \\
\hline
$(210)$ & 6.08  & $316.0^{+86.2}_{-69.3}$ & $1126.6^{+327.8}_{-214.9}$ & $290.5^{+71.1}_{-77.2}$ & $\times$ & $\times$  \\
$(220)$ & 2.95  & $217.8^{+80.6}_{-79.3}$ & $1125.5^{+190.6}_{-185.9}$ & $174.7^{+86.5}_{-77.3}$   & $3862.7^{+1466.1}_{-1003.3}$ & $97.6^{+37.9}_{-35.5}$ \\
$(320)$ & 3.06  & $124.5^{+130.4}_{-164.7}$ & $1105.8^{+191.9}_{-132.5}$ & $250.8^{+62.0}_{-40.2}$ & $C_\frac{3}{2}=C_\frac{1}{2}$ & $82.1^{+292.1}_ {-128.0}$\\ 
\hline
    \end{tabular}
\end{table}

 Reference~\cite{Du:2021bgb} investigated the dynamical impact of the $\Xi_c'\bar{D}$ (which in $S$ wave only interacts with $J=1/2$) on top of   $\jpsi\Lambda$ and $\Xi_c\bar{D}^*$. We number the channels according to their increasing thresholds, such that $\jpsi\Lambda$~(1), $\Xi_c'\bar{D}$~(2), and $\Xi_c\bar{D}^*$~(3). However, in order to avoid having too many free parameters this extra channel is introduced perturbatively, and its interactions would take place only through its coupling to $\Xi_c\bar{D}^*$. 
The resulting fit is (320) in the last panel of Fig.~\ref{fig.221023.1}. Its quality is good and it is also found that  the HQSS equality $C_{1/2}=C_{3/2}$ can be fulfilled. Then we conclude that the previously commented violation of this HQSS expectation in the fit (220) is due to having neglected a nearby threshold. 

We give the pole content of the fit (320) in the Table~\ref{tab.221023.6}, with the different RSs denoted as  RSII $(-,+,+)$, and RSIII $(-,-,+)$ (and its reduction for the two-channel case).
\begin{table}
\centering
\caption{\small The poles for the fit $(320)$ with three channels. \label{tab.221023.6} }
\begin{tabular}{lllllll}
  \hline
  Type & $J$ & RS    &  $\sqrt{s_R}$ & $|g_1|$ & $|g_2|$ & $|g_3|$ \\
         &     &       &   (MeV)       &  (MeV)     &  (MeV)  & (MeV)  \\
    \hline
$(320)$ & 3/2 & RSII $(-+)$   &  $4466.6^{+ 1.9}_{-2.7}-i\,1.3^{+1.3}_{-3.7}$ &  $1.4^{+1.4}_{-1.4}$  & $\times$ & $12.6^{+0.8}_{-0.6}$\\
$(320)$ & 1/2 & RSIII $(--+)$  &  $4453.8^{+ 2.4}_{-3.3}-i\,2.8^{+0.9}_{-0.8}$ &  $0.6^{+0.6}_{-0.6} $ &  $4.2^{+0.2}_{-0.4}$ & $15.0^{+0.5}_{-0.3}$  \\
    \hline
  \end{tabular}
  \end{table}
Having the residue and the pole positions we can calculate the partial-decay widths and partial compositeness coefficients, 
\begin{align}
\label{221023.18}
J=1/2:~& \Gamma_1= 0.5_{-0.5}^{+1.9}~{\rm MeV}\,, ~ \Gamma_2= 4.3 _{-1.4}^{+1.2}~{\rm MeV}\,, ~ \Gamma_3= 0.9_{-0.6}^{+1.2}~{\rm MeV}~,\\
&X_1= 0.0\pm 0.0\,, ~  X_2= 0.15\pm 0.05~,\nn\\
J=3/2:~& \Gamma_1= 2.6^{+8.2}_{-2.6}~{\rm MeV}\,, ~ \Gamma_3= 0.4 _{-0.4}^{+2.5}~{\rm MeV}~,~X_1= 0.0\pm 0.0\,, ~ X_3= 1.0^{+0.2}_{-0.2}~.\nn
\end{align}
From here one can conclude that the $P_{cs}$ is mostly a $\Xi_c\bar{D}^*$ resonance, having very large couplings to this channel driving to large $X_3$. The latter was not calculated for the $J=1/2$ pole since the threshold for the $\Xi_c\bar{D}^*$ channel lies clearly above its mass. Nonetheless,  the naively calculated $X_3=0.94$ in this case is very large, being a clear indication of the preponderance of the (3) channel in this pole. A similar comment is in order also for the $J=3/2$ pole, with $X_3$ having a central value of 1. This composite nature of the $P_{cs}(4459)$ is analogous to the one already unveiled for its non-strange pentaquark partners $P_c(4312)$, $P_c(4380)$, $P_c(4440)$, and $P_c(4457)$ \cite{Guo:2019kdc,Du:2019pij,Du:2021fmf}.


     \section*{Acknowledgements} 
{ This work has been supported in part by the MICINN
AEI (Spain) Grant No. PID2019–106080GB-C22/AEI/10.13039/501100011033\,, and the Natural Science Foundation of China (NSFC) under Grant No.~11975090.}

\bibliography{references}
\end{document}